\documentclass[11pt,tightenlines,eqsecnum,floats,aps,amsmath,amssymb,nofootinbib,prd,shownopacs,floatfix]{revtex4}

\usepackage{graphicx}
\usepackage{epstopdf}
\usepackage{latexsym}
\usepackage{amssymb}
\usepackage{amsmath}
\usepackage{color}
\usepackage{mathrsfs}
\usepackage{xparse}
\usepackage{float}
\usepackage{mathtools}
\usepackage{multirow}
\usepackage[center]{subfigure}
\DeclareMathOperator{\sinc}{sinc}

\begin{document}

  \renewcommand\arraystretch{2}
 \newcommand{\bq}{\begin{equation}}
  \newcommand{\lb}{\label}
 \newcommand{\eq}{\end{equation}}
 \newcommand{\bqn}{\begin{eqnarray}}
 \newcommand{\eqn}{\end{eqnarray}}
 \newcommand{\nb}{\nonumber}
 \newcommand{\cb}{\color{blue}}
    \newcommand{\cc}{\color{cyan}}
        \newcommand{\cm}{\color{magenta}}
\newcommand{\rc}{\rho^{\scriptscriptstyle{\mathrm{I}}}_c}
\newcommand{\rd}{\rho^{\scriptscriptstyle{\mathrm{II}}}_c} 
\NewDocumentCommand{\evalat}{sO{\big}mm}{%
  \IfBooleanTF{#1}
   {\mleft. #3 \mright|_{#4}}
   {#3#2|_{#4}}%
}
\newcommand{\PRL}{Phys. Rev. Lett.}
\newcommand{\PL}{Phys. Lett.}
\newcommand{\PR}{Phys. Rev.}
\newcommand{\CQG}{Class. Quantum Grav.}
\newcommand{\parallelsum}{\mathbin{\!/\mkern-5mu/\!}}
\title{Non-singular quantum gravitational dynamics of an LTB dust shell model: \\ the role of quantization prescriptions}
\author{Kristina Giesel$^{1}$}
\email{kristina.giesel@gravity.fau.de}
\author{Bao-Fei Li $^{2}$}
\email{baofeili1@lsu.edu}
\author{Parampreet Singh$^2$}
\email{psingh@lsu.edu}
\affiliation{$^{1}$ Institute for Quantum Gravity,  Department of Physics, FAU Erlangen-N\"urnberg, Staudtstr. 7, 91058 Erlangen, Germany\\
$^{2}$ Department of Physics and Astronomy, Louisiana State University, Baton Rouge, LA 70803, USA}

\begin{abstract}
We study some consequences of the loop quantization of the outermost dust shell in the Lema\^itre–Tolman–Bondi  spacetime with a homogeneous dust density using different quantization strategies motivated by loop quantum gravity. Prior work has dealt with loop quantizing this model by employing holonomies and the triads, following the procedure in standard loop quantum cosmology. In this work we compare this quantization with the one in which holonomies and gauge-covariant fluxes are used. While both of the quantization schemes resolve the central singularity, they lead to different mass gaps at which a trapped surface forms. This trapped surface which is matched to an exterior generalized Vaidya spacetime disappears when the density of the dust shell is in the Planck regime. We find that the  quantization based on holonomies and gauge-covariant fluxes generically results in an asymmetric evolution of the dust shell in which the effective mass associated with the white hole as seen by an external observer is $2/\pi$ of the one for the black hole. This effective difference in masses results from difference in the classical limits in pre- and post-bounce regimes in the two quantizations. This distinctive feature rules out formation of any black hole-white  hole twin in presence of gauge-covariant flux modifications which is in contrast to the  quantization using holonomies and triads where the gravitational collapse always leads to a black hole-white hole twins. In another striking difference, for the quantization based on holonomies and gauge-covariant  fluxes there can be situations in which during a non-singular collapse only a black hole forms without a white hole. 

\end{abstract}
\maketitle
\section{Introduction}

It is generally believed that the  singularity problem in the classical gravitational collapse is a consequence of  the breakdown of the Einstein theory of general relativity (GR) in the Planckian curvature regime and thus can be resolved by quantum gravity. Understanding the role of quantum gravity effects becomes manageable when the collapsing astrophysical body is assumed to be spherically symmetric in which case techniques from canonical quantum gravity can be applied to explore the consequences for singularity resolution  and the fate of the spacetime beyond the central singularity. It is expected that a rigorous understanding of quantum gravity effects would also yield insights on fundamental questions related to cosmic censorship conjecture and the black hole evaporation. But to answer these questions, it is important to reliably understand whether the resulting physics is tied to a particular quantization prescription and the way various quantization ambiguities affect the resulting physics of the quantum spacetime and the end state of gravitational collapse.

Loop quantum gravity (LQG) \cite{Rovelli_1998,ashtekar-lewandowski04,thiemann2001} is a non-perturbative and background independent approach of quantum gravity which provides a  platform for investigating the resolution of the singularities in various situations. In particular, when symmetry reduction is performed before quantization is carried out, loop quantization can in general lead to tractable models which grasp the main features of the quantum gravity corrections to the classical model. For example, a quantization of Friedmann-Lema\^itre-Robertson-Walker (FLRW) cosmological models using LQG techniques results in loop quantum cosmology (LQC) \cite{ashtekar-singh11,agullo-singh17} which resolves the big bang singularity replacing it with a big bounce at Planck curvature scale \cite{aps2006a,aps3,acs2010}. While the Hamiltonian constraint in LQC is a difference equation, the underlying dynamics is captured very well by an effective Hamiltonian which indicates a generic resolution of cosmological singularities for isotropic and  anisotropic spacetimes \cite{Singh_2009,Singh_2011,Singh:2011gp, Saini_2016,Saini_2017, saini_2018}. These studies have also been generalized to understand the way above results are robust to some underlying quantization ambiguities such as different choices of the Hamiltonian constraint \cite{mlqc-saini}.  

Similar to the cosmological setting, within the framework of symmetry reduced LQG, quantization of black hole spacetimes has been studied in various models 
 using effective spacetime description. For the vacuum spacetimes,  most studies are based on the fact that the interior of a Schwarzschild black hole is isometric to the vacuum Kantowski-Sachs spacetime in cosmology. Various quantization schemes, based on holonomies and symmetry reduced triads, have  been proposed which result in  a generic resolution of the  curvature singularity  and glue the black hole spacetime to a white hole spacetime through a transition surface  \cite{Ashtekar_2005, modesto_2005, Bohmer_2007, Campiglia_2008,Corichi_2016, Olmedo:2017lvt,  Ashtekar:2018lag, Ashtekar_2018, Bodendorfer_2019, han2020improved, Gan_2020}. Moreover, studies on the vacuum spacetimes have also be extended to include both interior and exterior of the black hole spacetimes \cite{Bojowald_2006,Gambini_2008,chiou2013loop,Gambini_2013,Gambini2014,Gambini_2020,Kelly_2020}.  On the other hand, investigations have been carried out beyond vacuum spacetimes to include matter such as a massless scalar field (see for eg.  \cite{Bojowald-singh2005,Goswami_2006,Gambini_2009,Bambi_2013,Tavakoli_2014,Ben_tez_2020}).  In the classical theory, one of the most studied context is of the dust collapse in Lema\^itre-Tolman-Bondi (LTB) spacetime \cite{lematre,tolman,bondi}, see for instance  \cite{gtt2010} for an analysis in terms of Dirac observables and 
 \cite{Bambi_2013,Bojowald:2008ja,wilson-ewing} where loop quantum gravitational effects have also been studied. Here the marginally bound case has been well studied in which the interior spacetime is isometric to a spatially-flat FLRW spacetime for a homogeneous evolution of the dust cloud. These studies have so far focused on exploring the resolution of central singularity using standard techniques in LQC by incorporating quantum geometry effects via holonomies and/or inverse triad modifications \cite{Bambi_2013,Bojowald_2009,wilson-ewing}. 
 
 Despite this progress,  none of the quantizations studied so far for symmetric models has been derived from LQG. Therefore, the robustness of physical predictions from above symmetric models when further modifications from LQG are included in the dynamics remains an important open question. One prominent approach to relate these models to LQG is via  coherent states to approximate certain sectors of LQG. But, this is challenging for quantizations, as ones discussed above, which are based on a fixed discretized lattice and flux variables  (for a discussion see  \cite{liegener-singh3,liegener-singh1,liegener-singh2}). A resolution is to use gauge-covariant flux variables on a discretized fixed lattice \cite{Thiemann_2001}. In this manuscript our goal is to test the robustness of some of the results obtained in the context of gravitational collapse of the dust cloud using this input from LQG and compare with existing approach based on using holonomies and triads by employing the $\bar \mu$ type quantization \cite{aps3}, first for holonomy and triads \cite{aps3}, and then holonomy and gauge-covariant fluxes. This allows to test some properties of gauge-covariant fluxes in a simpler setup than full LQG where detailed investigations on models involving gauge-covariant fluxes are still an open question. 
 We consider the collapse of a spherically symmetric dust cloud in the  marginally bound case assuming that  the energy density of the dust cloud is homogeneous so that each shell of the dust cloud collapses at the same relative velocity. As a result, the crossing of the dust shell would never occur during the collapse of the dust cloud and  the only relevant singularity in the classical theory is the central singularity when the radius of the dust cloud vanishes. 
 
 In this framework we investigate whether different quantization schemes can affect the formation of the trapped horizons and thus  lead to distinctive phenomenological signals once the black hole forms during the collapse of the dust cloud. From the numerical simulations of the effective Hamilton's equations, we find the central singularity is resolved and replaced by a quantum bounce within  both quantization schemes. However, there are also distinctive features arising in the second scheme due to the asymmetric evolution of the dust shell. For example, there exists a small region of the parameter space in which a black hole can form during the collapse while a white hole in the expanding stage after the bounce cannot form. Besides, even when both the black hole and the white hole can form in the second scheme, their masses and the duration of the existence of trapped surfaces show qualitatively different behavior from the first scheme. In particular, for generic initial conditions the effective mass of the white hole as seen by an external observer turns out  to be  $2/\pi$ of the black hole mass. We find that in both the prescriptions there is a mass gap but its value depends on the chosen quantization scheme. These results provide insights on the role of the quantization prescription used in studying the fate of the gravitational collapse using LQG techniques. In particular our results demonstrate in a simple setting that while the resolution of central singularity is a robust feature, the existence of a ``black hole-white hole'' {\it{twin}} resulting from the Planck scale physics  is not a generic feature. It shows that the resulting physics of loop quantum gravitational collapse can be much richer and complex if further modifications from LQG, than those considered so far, are systematically incorporated.

This manuscript is organized as follows. In Sec. \ref{sec:shell model}, we briefly review the classical LTB dust shell model  and derive the classical Hamilton's equations of the outermost shell for the marginally bound case along the lines of \cite{Kiefer:2019csi} using Ashtekar-Barbero variables. We also discuss the criterion for the formation of the trapped surfaces during the collapse of a spherically symmetric object and the match of the interior of the trapped surface with the exterior spacetimes at the boundary.  In Sec. \ref{sec:effective dynamics}, we study the effective dynamics of the  loop quantization of the dust shell model and employ two quantization prescriptions, the first one only considers the holonomy corrections and the second takes into account both holonomy corrections and gauge covariant fluxes. With the help of the numerical analysis of the resulting  effective Hamilton's equations, we discuss the phenomenological differences due to the quantization prescriptions. Finally, the main results  are summarized  in Sec. \ref{sec:summary}. We use Planck units for numerical simulations and set $\hbar=c=1$ while keeping Newton's constant $G$ explicit in our formulae. 

\section{A brief review of the classical LTB dust shell model}
\label{sec:shell model}

In this section we briefly review  the classical LTB dust model in the canonical framework  and write necessary equations using triad and connection variables. 
The LTB model is obtained from  a spherical symmetric solution of Einstein's equations that involve non-rotational dust as the matter source. The metric is given by
\begin{equation}
\label{eq:metricLTB}
ds^2 = -d\tau^2 +\frac{(R^\prime)^2}{1+2f}dx^2+R^2d\Omega^2    
\end{equation}
where we denote the radial coordinate by $x$, the angular part by $d\Omega^2=d\theta^2+\sin^2\theta d\phi^2$ and we have set $c=1$. $R(x,\tau)$ which is the areal radius of the spherical surfaces  is determined by two unknown functions $F$ and $f$ that depend on the radial coordinate and are in turn determined via the following equations
\begin{equation}
\label{classical equations of motion}
8\pi G\rho_{\rm dust} =  \frac{F^\prime}{R^2R^\prime}\quad{\rm and}\quad
\dot{R}^2=\frac{F}{R}+2f,
\end{equation}
where $\rho_{\rm dust}$ denotes the energy density of the dust and $G$ is Newton's constant. Here $f(x)$ is the total energy of a unit mass at $x$ 
and $F(x)/2G$ is the active gravitating mass within a sphere with radius $R(x)$. 
In the so called marginally bound case one chooses $f=0$ leading to a simplified form of the metric as well as the LTB equations. We  consider the marginally bound case in this work.
In the Hamiltonian framework using canonical variables in the ADM formulation $(R,P_R)$, the Hamiltonian constraints for the gravitational  and dust matter  sectors have the form
\begin{equation}
\label{eq:ADMHam}
H_{\rm grav} = -\frac{1}{2G}\left(\dot{R}^2R\right)^\prime\quad{\rm and}\quad H_{\rm dust} = \frac{F^\prime}{2G},
\end{equation}
where $\dot{R}$ is understood as a function of the momentum $P_R$. Considering the total Hamiltonian constraint $H_{\rm grav} + H_{\rm dust}=0$ then yields the second equation in \eqref{classical equations of motion}.

In order to apply LQG techniques as used in LQC in this context we need to start with a classical LTB model formulated in terms of Ashtekar variables (see for instance \cite{Bojowald:2008ja}).  The set of independent canonical variables is given by $(A_x(x),E^x(x))$, $(\gamma K_\varphi(x),E^\varphi(x))$ and a third pair $(\eta(x),P_\eta(x))$ corresponding to  the gauge angle, where $\gamma$ is the Barbero-Immirzi parameter \cite{Barbero:1994ap,Immirzi:1996di}. Once the Gauss constraint is implemented the latter pair is eliminated and gauge invariant quantities do not depend on these variables. The relation between the connection variables $A_x$ and the corresponding extrinsic curvature $K_x$ in this case reads $K_x=\frac{1}{\gamma}A_x$. Therefore, the elementary canonical variables satisfy the following Poisson brackets
\begin{equation}
\{K_x(x),E^x(y)\}=2 G\,\delta(x,y)\quad{\rm and}\quad \{K_\varphi(x),E^\varphi(y)\}=G\,\delta(x,y) .  
\end{equation}
 The metric in the marginally bound case in terms of Ashtekar variables can be expressed as
\begin{equation}
ds^2  = -d\tau^2  +\frac{(E^\varphi)^2}{|E^x|}dx^2 +|E^x|d\Omega^2 .
\end{equation}
In the ADM case for spherical symmetric models one starts with two sets of variables $(R,P_R)$ and $(\Lambda,P_\Lambda)$. For the LTB marginally bound case, as can be seen in \eqref{eq:metricLTB}, the metric depends only on $R$ and its derivatives since 
$\Lambda=R^\prime$ \cite{gtt2010}. At the level of Ashtekar variables this results in the following LTB condition
\begin{equation}
\label{eq:LTBCond}
E^\varphi(x) = \frac{1}{2}|E^x|^\prime (x)\, .
\end{equation}
As discussed in \cite{Bojowald:2008ja} a second condition relating the two extrinsic curvatures can be obtained by taking the LTB condition in (\ref{eq:LTBCond}) as a gauge fixing condition for the diffeomorphism constraint which then  yields
\begin{equation}
\label{eq:LTBCond2}
K^\prime_\varphi = K_x \, {\rm sgn}(E^x) .   
\end{equation}
The Hamiltonian constraint for the gravitational sector in terms of Ashtekar variables is given by
\begin{equation}
H_{\rm grav} = - \frac{1}{2G}\left(\frac{K_\varphi^2E^\varphi}{\sqrt{|E^x|}}+2K_{\varphi}K_x\sqrt{|E^x|} \right),   
\end{equation}
where at this stage $K_\varphi,E^\varphi$ are understood as functionals of $K_x,E^x$. In order to write down $H_{\rm grav}$ entirely in terms of $K_x,E_x$ one can use the equations of motions for $K_\varphi,E^\varphi,K_x,E^x$ to get \cite{Bojowald:2008ja}
\begin{equation}
{K}_\varphi= \frac{\dot{E}^x}{2\sqrt{|E^x|}}\quad{\rm and}\quad {K}_x = \frac{1}{\sqrt{|E^x|}}\left( \dot{E}^\varphi - \frac{\dot{E}^xE^\varphi}{2E^x}\right) .  
\end{equation}
If one reinserts these into $H_{\rm grav}$ one obtains 
\begin{equation}\label{Hameq}
H_{\rm grav} = -\frac{1}{2G}\left(\frac{(\dot{E}^x)^2}{4\sqrt{|E^x|}} \right)^\prime, 
\end{equation}
whereas the dust contribution has again the form $H_{\rm dust} = \frac{F^\prime}{2G}$. This is consistent with the ADM result \eqref{eq:ADMHam} if one considers the usual relation $R=\sqrt{|E^x|}$. 

Now we specialize the LTB model to the shell model. 
Our strategy is to consider the LTB conditions at the classical level which is sufficient to obtain a model for the outermost shell following the work in \cite{Kiefer:2019csi}. We will briefly summarize the main steps to obtain this model. One of the crucial ingredients of the model is that considering the LTB equations given by
\begin{equation}
\label{eq:EOMLTBAshtekar}
8\pi G\rho_{\rm dust} =  \frac{2F^\prime{\rm sgn}(E^x)}{\sqrt{|E^x|}(E^x)^\prime}\quad{\rm and}\quad
(\dot{E}^x)^2=4F\sqrt{|E^x|},
\end{equation}
where the second equation can, as before, be obtained by means of the total Hamiltonian constraint in Ashtekar variables using \eqref{Hameq}. One realizes that the equation for $\dot{E}^x$ only depends on $F$ and $E^x$ but not its spatial derivatives. As a consequence, at the classical level, once a mass function $F$ is chosen the individual shells decouple. The starting point for the outermost shell model in \cite{Kiefer:2019csi} is the Einstein-Hilbert action symmetry reduced to the LTB case plus a boundary term. Next, one notes that the trace of the Einstein-tensor in four dimensions yields $-{\cal R}(x)$, where ${\cal R}$ denotes the Ricci scalar. Multiplying the trace of the Einstein equations by a volume term then leads to
\begin{equation}
\sqrt{|g|}{\cal R}(x) = 8\pi\rho_{\rm dust}\frac{1}{2}{\rm sgn}(E^x)(E^x)^\prime\sqrt{|E^x|}\sin\theta = \frac{1}{G}F^\prime\sin\theta\, ,   
\end{equation}
where one uses the first LTB equation in \eqref{eq:EOMLTBAshtekar}. We denote the radial coordinate of the outermost shell by $x_0$, then after integrating out the angular part the bulk action has the form
\begin{equation}
S_{\cal M} = \frac{1}{4G}\int d\tau \int\limits_{0}^{x_0} F^\prime(x) = \frac{1}{4G}\int d\tau \left(F(x_0) - F(0)\right)
= \frac{1}{16G}\int d\tau\frac{(\dot{E}^x)^2(x_0)}{\sqrt{|E^x|(x_0)}}
\end{equation}
where as in \cite{Kiefer:2019csi} $F(0)=0$ is chosen, meaning that the innermost shell contains no mass. 
Let us introduce the following compact notation for the quantities of the outermost shell
\begin{equation}
\varepsilon^x \coloneqq E^x(x_0) .
\end{equation}
Using the same boundary term as derived in \cite{Kiefer:2019csi} which in this notation reads $S_{\cal B}=-\frac{3}{16G}\int d\tau \frac{(\dot{\varepsilon}^x)^2}{\sqrt{|\varepsilon^x|}}$ and adding the bulk and boundary contributions
finally, one  obtains the full action for the outermost shell of the dust cloud in the LTB model, which reads
\bq
\label{action for outermost shell}
S=\frac{1}{G}\int d\tau L_{\rm shell}\coloneqq -\frac{1}{8G}\int d\tau \frac{(\dot{\varepsilon}^x)^2}{\sqrt{|\varepsilon^x|}},
\eq
where as in \cite{Kiefer:2019csi}  it was used that for Brown-Kucha\v{r} dust \cite{Brown:1994py} the action trivially vanishes on-shell. 
Moreover, the triad $\varepsilon^x$ and its conjugate momentum $k_x$ which corresponds to the $1/2$ of the radial component of  the extrinsic curvature  satisfy the Poisson bracket 
\bq
\label{Poisson}
\{k_x,\varepsilon^x\}=G,
\eq 
then in terms of these canonical variables, the Hamiltonian of the outermost shell of the dust cloud takes the form
\bq
\label{Hamiltonian}
H=-\frac{2}{G} k_x^2\sqrt{|\varepsilon^x|}=-M,
\eq
where $M$ stands for the dust mass enclosed by the outermost dust shell. 
 The classical Hamilton's equations can be easily derived from the Hamiltonian with the above Poisson bracket, which explicitly read
\bq
\label{classical-eom}
\dot \varepsilon^x=4k_x\sqrt{|\varepsilon^x|},\quad \quad \dot k_x=-\frac{k^2_x}{\sqrt{|\varepsilon^x|}}.
\eq

 We can rewrite $\varepsilon^x$ again in terms of $R(x_0)=R_0$  as $|\varepsilon^x| = R_0^2$ where we suppress the label ``0" from now on. The above equations of motion in turn yield the following equations which resemble the classical Friedmann and the Raychaudhuri equations for the areal radius
 \bq
\label{classical-FR}
\left(\frac{\dot R}{R}\right)^2=\frac{8\pi G}{3}\rho,\quad \quad \frac{\ddot R}{R}=-\frac{4\pi G}{3}\rho .
\eq
Here $\rho$ denotes the energy density of the dust cloud 
\bq
\rho=\frac{3M}{4\pi R^3} ~.
\eq

As is well-known, the central singularity in the classical theory is inevitable since dynamical equations (\ref{classical-FR}) result in the radius of the outermost dust shell to decrease to zero in a finite period of the proper time for a generic set of the initial conditions $(R_i,M)$, where $R_i$ denotes the initial value of the radius. During the collapse of the dust cloud, if a trapped surface can form, then the central singularity will be covered by a horizon. In order to investigate the formation of the trapped surfaces in the interior, it is convenient to introduce two future-directed null vectors normal to the sphere with constant radius, which are \cite{hayward96}
\bq
\label{future-pointed}
\partial_{\xi^+}=\frac{1}{\sqrt 2}\left(\partial_\tau+\frac{1}{R'}\partial_x\right),\quad \partial_{\xi^-}=\frac{1}{\sqrt 2}\left(\partial_\tau-\frac{1}{R'}\partial_x\right). 
\eq
In the null coordinates $\xi^+$ and $\xi^-$, the line element (\ref{eq:metricLTB}) (with $f=0$) becomes
\bq
ds^2=-2d\xi^+d\xi^-+R^2d\Omega^2.
\eq
Hence, we can identify two kinds of the radial null geodesics emerging from the sphere, namely the inward null geodesics $\xi^+=const$ and the outward null geodesics $\xi^-=const$. If the radius decreases along both inward and outward null geodesics, then a trapped surface forms at the sphere. One can introduce the expansion parameter \cite{hayward96}
\bq
\theta_{\pm}=\frac{2}{R}\partial_{\pm}R,
\eq
where $\partial_{\pm}$ denotes derivatives with respect to $\xi^{\pm}$ respectively.  When the bundles of the light rays converge on both sides of the sphere, $\theta_{\pm}<0$, namely $\dot R<-1$, then the sphere becomes a future trapped surface corresponding to a black hole. When $\theta_{+}\theta_{-}=0$, the sphere is marginally trapped. On the other hand, if we reverse the directions of the null vectors in (\ref{future-pointed}), and then require the bundles of the light rays converge along these reversed null vectors, which leads to $\dot R>1$, a past trapped surface corresponding to a white hole can form. As a result, in both cases,  the trapped surfaces form as long as $\dot R^2>1$ with the sign of $\dot R$ signifying future ($\dot R < 0$) or past ($\dot R>0$) directed trapped surface.

While the simplest models of the gravitational collapse, such as Oppenheimer-Snyder-Dutt model, can be recasted as a cosmological model it should be noted that this analogy is only true in the interior of the collapsing object. For a complete picture of a gravitational collapse  the interior has to be matched with an exterior spacetime  for an external observer. In various situations, with a non-vanishing interior pressure, including for the analysis presented in this manuscript, the exterior spacetime turns out to be the generalized Vaidya spacetime \cite{wang-wu99,joshi-dwivedi99}, which in the advanced Eddington-Finkelstein coordinates takes the form 
\bq
\label{Vaidya}
ds^2_+=-\left(1-\frac{2GM(r_v,v)}{r_v}\right)dv^2-2dvdr_v+r^2_vd\Omega^2 .
\eq
Here $M(r_v,v)$ is the generic mass function of the Vaidya spacetime which stands for the mass of the black hole if a trapped surface can form during the collapse of the dust cloud. The interior and exterior metrics (\ref{eq:metricLTB}) (with $f=0$) and (\ref{Vaidya}) can be matched at the outermost dust shell $x=r_b$ by requiring that the first and second fundamental forms are equal at the boundary. This leads to the following equations
\bqn
\label{matching conditions}
r_v(v)|_\Sigma&=&R(r_b,t)=r_b a(t),\quad \quad \left(\frac{dv}{dt}\right)|_\Sigma=\frac{R_{,x}+r_b\dot a}{1-F/R},\\
\label{matching conditions 2}
F(t,r_b)&=&2M(r_v,v)G,\quad  \quad  M(r_v,v)_{,r_v}G=\frac{F}{2R}+r^2_ba\ddot a.
\eqn

\section{Effective dynamics of the loop quantized LTB dust shell model}
\label{sec:effective dynamics}

In this section, we focus on the evolution of the outermost shell of the dust cloud when the quantum geometric effects motivated from LQG are taken into consideration. Using the techniques of the effective description of the loop quantum dynamics,  we loop quantize the outermost shell of the dust cloud and present two effective Hamiltonians resulting from two loop quantizations which yield different physical results.  These effective Hamiltonians  incorporate the quantum effects at the Planckian energy density and thus successfully resolve the central singularity encountered during the collapse of the  classical dust cloud. We then present numerical results from quantum gravity modified dynamical equations and discuss the way initial conditions determine the formation of the trapped horizons during the evolution of the dust shell.

\subsection{Quantum gravity modified dynamical equations}

In this subsection, we consider effective dynamics resulting from two loop quantizations of  the classical  Hamiltonian (\ref{Hamiltonian}). The first is based on using holonomy and triad variables as in standard LQC \cite{abl,aps3}, and the second is based on a recently studied quantization of holonomy and gauge-covariant fluxes \cite{liegener-singh1,liegener-singh2,liegener-singh3}. In addition to the effective Hamiltonian  and the equations of motion in each case, we also discuss analytical solutions in the first case and define useful variables for unravelling the dynamics of the collapsing dust cloud through numerical results when analytical solutions are not available. 

\subsubsection{Non-singular evolution in holonomy and triad quantization}
\label{sec:holonomy corrections}
The effective description of the quantum dynamics with the holonomy corrections  has been widely corroborated with numerical simulations for both isotropic and anisotropic models in the context of LQC. The modified dynamical equations from the effective Hamiltonian turn out to be an excellent approximation to the underlying quantum dynamics for states which are sharply peaked in a large macroscopic universe  \cite{numlsu-2,numlsu-3,numlsu-4,numlsu-1}. Following these investigations, we assume the validity of this approach in our analysis. The marginally bound case corresponds to a spatially-flat spacetime in a cosmological setting for which the Ashtekar-Barbero connection is related to extrinsic curvature by a multiplicative constant and the holonomy corrections can be incorporated into an effective Hamiltonian by making the substitution $k_x\rightarrow \sin(\delta_x k_x)/\delta_x$ in the classical Hamiltonian ($\ref{Hamiltonian}$).   In this way, the quantum geometric effects are manifest in the regime where $\delta_x k_x\gg1$ and the effective Hamiltonian approximates the classical one when $\delta_x k_x\ll1$. Here one needs to be careful with the way $\delta_x$ depends  on or is independent of the phase space variables, otherwise one can get inconsistent infra-red regime and fake Planck scale effects. In the spatially-flat cosmological setting, there is a unique prescription which is shown to be viable \cite{cs08}, known as improved dynamics \cite{aps3}. The same prescription has been shown to be a viable one for the non-marginally bound case too \cite{bfl-ps}. In this quantization prescription, the dependence of $\delta_x$ on the triads is computed by equating the physical area enclosed by honolomy to the minimal eigenvalue of the area operator in LQG which is $\Delta= 4\sqrt{3}\pi \gamma l_{pl}^2$ \cite{aps3}. In the following, the Barbero-Immirzi parameter  $\gamma$ is chosen to be $0.2375$ which is fixed by  the black hole thermodynamics in LQG. To find this relationship we use the correspondence with the cosmological model and note that 
\bq
k_x=\frac{c}{2\gamma}\left(\frac{3}{4\pi}\right)^{1/3},\quad \quad \varepsilon^x=p\left(\frac{3}{4\pi}\right)^{2/3}.
\eq
This sets $\delta_x=2\gamma \lambda/\sqrt{\varepsilon^x}$ with $\lambda=\sqrt \Delta$,
here we have taken the positive orientation of the triad. 
As a result, the effective Hamiltonian  from the holonomy corrections takes the form
\bq
\label{ham1}
H^{\scriptscriptstyle{\mathrm{hol}}}_\mathrm{eff}=-\frac{ (\varepsilon^x)^{3/2} }{2 G\lambda^2\gamma^2} \sin^2\left(\frac{2 \gamma \lambda k_x}{\sqrt{\varepsilon^x}}\right)=-M,
\eq
where  $k_x/\sqrt{\varepsilon^x}$ is proportional to the relative velocity $\dot R/R$ of the dust shell in the classical limit. From the effective Hamiltonian (\ref{ham1}), it is straightforward to show that the equations of motion are 
\bqn
\label{3a1}
\dot \varepsilon^x&=& \frac{\varepsilon^x}{\gamma \lambda}\sin\left(\frac{4\lambda \gamma k_x}{\sqrt{\varepsilon^x}}\right),\\
\label{3a2}
\dot k_x&=&\frac{k_x}{2\lambda\gamma}\sin\left(\frac{4\lambda \gamma k_x}{\sqrt{\varepsilon^x}}\right)-\frac{3\sqrt{\varepsilon^x}}{4\gamma^2\lambda^2}\sin^2\left(\frac{2\lambda \gamma k_x}{\sqrt{\varepsilon^x}}\right).
\eqn
From the equation of motion (\ref{3a1}), one can find the dynamical equation:
\bq
\label{velocitysquare}
\left(\frac{\dot R}{R}\right)^2 = \frac{8\pi G}{3}\rho\left(1-\frac{\rho}{\rho^{\scriptscriptstyle{\mathrm{hol}}}_\mathrm{max}}\right),
\eq
where $\rho=\frac{3M}{4\pi R^3}$ and $\rho^{\scriptscriptstyle{\mathrm{hol}}}_\mathrm{max}=3/(8\pi G\lambda^2\gamma^2)$ denotes the maximum energy density enclosed by the outermost dust shell that is allowed in this model. We see that the quantum geometric effects modify the classical term $8\pi G \rho/3$ with an additional factor which vanishes when density of the dust cloud  becomes equal to $\rho^{\scriptscriptstyle{\mathrm{hol}}}_\mathrm{max}$. Hence, the radius of the dust shell attains its minimum value at the highest energy density and consequently the singularity encountered at $R=0$ is avoided. Note that the above modified Friedmann equation results in the classical Friedmann equation (\ref{classical-FR}) at macroscopic scales both in the pre-bounce and post-bounce phases. 

The quantum gravitational modification in (\ref{velocitysquare}) can be understood in terms of an ``effective energy density'' by writing its right hand side in the same form as in the classical theory. That is by defining 
\bq
\label{3a3}
\rho^{\scriptscriptstyle{\mathrm{hol}}}_\mathrm{eff}\coloneqq \rho\left(1-\frac{\rho}{\rho^{\scriptscriptstyle{\mathrm{hol}}}_\mathrm{max}}\right),
\eq
such that 
\bq\label{velocitysquare2}
\left(\frac{\dot R}{R}\right)^2= \frac{8\pi G}{3} \rho^{\scriptscriptstyle{\mathrm{hol}}}_\mathrm{eff} ~.
\eq
Using this effective energy density one can also define an effective mass 
\bq
\label{effective mass holonomy}
 M_\mathrm{eff}^{\scriptscriptstyle{\mathrm{hol}}}\coloneqq \frac{4\pi}{3}R^3 \rho_\mathrm{eff}^{\scriptscriptstyle{\mathrm{hol}}}.
 \eq
The physical meaning of this effective mass $M_\mathrm{eff}^{\scriptscriptstyle{\mathrm{hol}}}$ can be interpreted as follows. Note the Vaidya mass $M(r_v,v)$ is matched with the mass function at the outermost shell of the dust cloud as given in \eqref{matching conditions 2}, namely, $F=2GM(r_v,v)$. With $F=\dot R^2 R$, one can easily find from \eqref{velocitysquare2} and \eqref{effective mass holonomy} that $F=2 G M_\mathrm{eff}$. As a result, when the trapped surface forms during the collapse of the dust cloud, the effective mass is exactly the mass of the black hole observed by an outside spectator. Same arguments can also be applied to the case in which a white hole forms in the expanding phase of the dust cloud after the bounce. There, the effective mass stands for the mass of the white hole as detected by the exterior observer.   Finally, the effective energy density obeys a conservation law  
\bq
\dot {\rho}_\mathrm{eff}^{\scriptscriptstyle{\mathrm{hol}}}+3\frac{\dot R}{R} \left(\rho_\mathrm{eff}^{\scriptscriptstyle{\mathrm{hol}}}+p_\mathrm{eff}^{\scriptscriptstyle{\mathrm{hol}}}\right)=0.
\eq
 with an effective pressure 
\bq
p_\mathrm{eff}^{\scriptscriptstyle{\mathrm{hol}}}=-\frac{\dot M_\mathrm{eff}^{\scriptscriptstyle{\mathrm{hol}}}}{4\pi R^2\dot R}~.
\eq

The formation of the trapped surface depends on the magnitude of $\dot R^2$ during the collapse of the dust cloud. Specifically, the trapped surface would form if and only if the magnitude of $\dot R^2$ becomes greater than unity. This results in a threshold value of the dust mass $M^*$ below which  trapped surface does not form. Only when $M>M^*$, a trapped surface would  form.  On the other hand, if $M<M^*$, no trapped region forms during the collapse of the dust shell. 

For the physical Hamiltonian (\ref{ham1}), 
one can analytically find the exact expression of $M^*$. First, from (\ref{velocitysquare}), it is straightforward to show that 
\bq
\lb{velocity-squared}
\dot R^2=\frac{8\pi G M^{2/3}\rho^{1/3}}{(48\pi^2)^{1/3}}\left(1-\frac{\rho}{\rho^{\scriptscriptstyle{\mathrm{hol}}}_\mathrm{max}}\right).
\eq
Therefore, at $\rho=\rho^{\scriptscriptstyle{\scriptscriptstyle{\mathrm{hol}}}}_\mathrm{max}/4$, $\dot R^2$ attains its maximum value which turns out to be 
\bq
\label{3a4}
\dot R^2_\mathrm{max}=\frac{3}{4}\left(\frac{GM}{\lambda\gamma }\right)^{2/3}.
\eq
Requiring $\dot R^2_\mathrm{max}=1$ yields the threshold value of the dust mass which is 
\bq
\label{threshold}
M^*=\frac{8\lambda\gamma}{3\sqrt 3 G}\approx0.831,
\eq
with $\gamma=0.2375$ and $\lambda\approx 2.2736$ in Planck units. Moreover, since $\dot R^2$ given by (\ref{velocity-squared}) only depends on the dust mass and the energy density of the dust cloud, the energy density at which the trapped surface forms or vanishes also only depends on the dust mass. In particular, the initial volume does not play any role in determining whether the trapped surface would form during the collapse of the dust cloud.

In addition to the threshold value of the dust mass for the formation of the trapped surface, one can also analytically obtain the time at which trapped surface forms. Integrating (\ref{3a3}), leads to
\bq
\label{3a5}
R=\Big[2MG\lambda^2\gamma^2+\left(\alpha \tau+\beta\right)^2\Big]^{1/3},
\eq
where
\bq
\label{3a6}
\alpha=-3\sqrt{GM/2},\quad \quad \beta=\sqrt{R^3_i-2MG\lambda^2\gamma^2},
\eq
and $R_i$ denotes the initial value of the radius $R$ with the initial time chosen at $\tau_i=0$. Thus, the bounce happens at $\dot R=0$ when 
\bq
\label{bouncetime}
\tau_B=\frac{2\sqrt{R^3_i-2MG\lambda^2}}{3\sqrt{2GM}}.
\eq
More features of the dust cloud collapse with holonomy corrections will be discussed through the numerical solutions in the next subsection.\\

\subsubsection{Non-singular evolution in holonomy and gauge-covariant flux quantization}
\label{sec:gauge covariant flux}
So far we have considered loop quantum gravity effects using holonomies of the connection and {\it symmetry reduced triads}. The usage of triads, instead of fluxes, is possible because of symmetry reduction. While this strategy works for loop quantization of symmetry reduced models, one needs to go beyond this approach if one wishes to obtain an effective Hamiltonian with loop quantum modifications from loop quantum gravity using suitable coherent states. A possibility in this direction requires an introduction of {\it gauge-covariant fluxes}, first introduced by Thiemann \cite{Thiemann_2001}, which have been recently implemented in loop quantization of cosmological spacetimes \cite{liegener-singh1,liegener-singh2,liegener-singh3}. It tuns out that the corresponding quantum effects can be incorporated into the effective Hamiltonian by making the substitution $\sqrt {\varepsilon^x}\rightarrow \sqrt{\varepsilon^x}\sinc\left(\delta_x k_x/2\right)$ in the classical Hamiltonian (\ref{Hamiltonian}), which, together with the holonomy corrections in $k_x$, gives rise to the following effective Hamiltonian, 
\bq
\label{ham2}
H^{\scriptscriptstyle{\mathrm{g.c.}}}_\mathrm{eff}=-\frac{(\varepsilon^x)^{3/2}}{2 G\gamma^2 \lambda^2} \sin^2\left( \frac{2 \lambda \gamma k_x}{\sqrt{\varepsilon^x}}\right)\sinc\left(\frac{\gamma \lambda k_x}{\sqrt{\varepsilon^x}}\right)=-M.
\eq
To distinguish this effective Hamiltonian from the one where only holonomy modifications were incorporated, we label it with superscript ``g.c." for inclusion of gauge-covariant flux modifications in addition to holonomy modifications.

Correspondingly, the equations of motion in this case are given by 
\bqn
\label{eom1}
\dot \varepsilon^x&=&\frac{(\varepsilon^x)^{3/2}}{2\lambda^2\gamma^2k_x}\sin\left(\frac{2\gamma \lambda k_x}{\sqrt{\varepsilon^x}}\right)\sin\left(\frac{\lambda\gamma k_x}{\sqrt{\varepsilon^x}}\right)\bigg\{1+5\cos\left(\frac{2\gamma \lambda k_x}{\sqrt{\varepsilon^x}}\right)-\frac{\sqrt{\varepsilon^x}}{\gamma \lambda k_x}\sin\left(\frac{2\gamma \lambda k_x}{\sqrt{\varepsilon^x}}\right)\Bigg\},\\
\label{eom2}
\dot k_x&=&\frac{\sqrt{\varepsilon^x}}{4\lambda^2\gamma^2}\sin\left(\frac{2\gamma \lambda k_x}{\sqrt{\varepsilon^x}}\right)\sin\left(\frac{\gamma \lambda k_x}{\sqrt{\varepsilon^x}}\right)\bigg\{1+5\cos\left(\frac{2\gamma \lambda k_x}{\sqrt{\varepsilon^x}}\right)-4\frac{\sqrt{\varepsilon^x}}{\lambda\gamma k_x}\sin\left(\frac{2\gamma\lambda k_x}{\sqrt{\varepsilon^x}}\right)\Bigg\}.
\eqn
The classical equations of motion (\ref{classical-eom}) can be recovered from the above equations in the limit when $k_x/\sqrt{\varepsilon^x}\ll1$. While the above dynamical equations result in a non-singular bounce as in standard LQC, there are some key differences in the evolution. In particular, in presence of gauge-covariant fluxes matter acts non-minimally coupled and the bounce turns out to be generically asymmetric \cite{liegener-singh1,liegener-singh2}.

In the case with the gauge covariant fluxes,  a closed form of the modified Friedmann equation analogous to (\ref{3a3}) is not available.   As a result, an analytical analysis of the the threshold value of the dust mass for the formation of the trapped horizon is not possible. However, one can use the numerical simulations to understand the dynamics of the collapse of the dust cloud in this model by using the criteria for the formation of a trapped surface which is still given by $\dot R^2\ge1$. Besides, the asymptotic form of the Friedmann equation can be obtained from the large volume approximation as what has been done for a massless scalar field in \cite{liegener-singh1}. Due to the non-minimal coupling of the matter sector with gravity, the asymptotic form  of the Friedmann equation also changes with the equation of state of the matter content. For the dust field, following the similar calculations in \cite{liegener-singh1}, it is straightforward to show that in the collapsing phase of the dust cloud, the asymptotic form of the Friedmann equation assumes
\bq
\label{Friedmann-collapse}
H^2|_\mathrm{collapse}=\frac{8\pi G}{3}\rho+\mathcal O\left(\rho^2\right),
\eq
while the asymptotic form of the Friedmann equation in the expanding phase of the dust cloud reads
\bq
\label{Friedmann-expanding}
H^2|_\mathrm{expanding}=\frac{8\pi \alpha G}{3}\rho+\mathcal O\left(\rho^{3/2}\right),
\eq
with $\alpha =2/\pi$.\footnote{This additional factor $\alpha$ can also be absorbed into the Newton's constant, yielding a rescaled Newton's constant $\tilde G=\alpha G$.} The origin of this $\alpha$ in holonomy-gauge-covariant flux modifications is tied to the "$\mathrm{sinc}$" term in the Hamiltonian. Note that in both quantizations, the post-bounce classical branch corresponds to $ |2 \gamma  k_x/\sqrt{\varepsilon^x}|$ reaching $\pi/\lambda$. In standard quantization based on holonomies and triads this is indistinguishable from the pre-bounce classical branch $2 \gamma k_x/ \sqrt{\varepsilon^x} \approx 0$. The pre-bounce and post-bounce branches in this quantization are thus identical in the classical limit. However, the $\mathrm{sinc}$ term in the presence of gauge covariant flux modifications results in {\it{different}} classical limits in the pre-bounce and post-bounce epochs. These branches though still correspond to Friedmann dynamics, they have difference due to above $\alpha$ scaling which is the cause of asymmetry in effective mass of white hole as seen by an external observer.

Now following (\ref{velocitysquare2}), one can define the same effective energy density and the effective mass, namely,
\bq
\rho_\mathrm{eff}\coloneqq \frac{3}{8\pi G} \frac{\dot R^2}{R^2}, \quad \quad M_\mathrm{eff}\coloneqq \frac{4\pi}{3}R^3 \rho_\mathrm{eff}.
\eq
The effective energy density $\rho_\mathrm{eff}$ determines the relative collapsing (expanding) speed of the dust outermost shell, while the effective mass  $M_\mathrm{eff}$ which is the same as the Vaidya mass determines the mass of the black hole (during the collapsing phase) or the white hole (during the expanding phase) if  trapped surfaces are formed. When the volume is macroscopic and energy density is far below the Planck scale, using (\ref{Friedmann-collapse}), we find during the collapse of the dust cloud
\bq
\label{mass in contracting phase}
M_\mathrm{eff}|_\mathrm{collapse}=\frac{R^3}{2 G}H^2|_\mathrm{collapse}\approx\frac{4\pi}{3}R^3 \rho=M.
\eq
Therefore, initially the effective mass is equal to the mass of the dust cloud. On the other hand, after the bounce, when the dust cloud expands to a low energy density $\rho$($\ll\rho_\mathrm{pl}$), using (\ref{Friedmann-expanding}), one can easily find 
\bq
\label{mass in expanding phase}
M_\mathrm{eff}|_\mathrm{expanding}=\frac{R^3}{2 G}H^2|_\mathrm{expanding}\approx\frac{4\pi \alpha}{3}R^3 \rho=\alpha M,
\eq
which implies that after the bounce, the effective mass of the dust cloud is just a fractional of the initial mass from the perspective of an outside observer. The same argument also applies to the case when the trapped surface forms, leading to a black hole and white hole asymmetry. 
Accompanied with the effective energy density, one can also define an effective pressure via
\bq
p_\mathrm{eff}=-\frac{\dot M_\mathrm{eff}}{4\pi R^2\dot R},
\eq
which satisfies
\bq
\lb{energy conservation in effective quantities}
\dot \rho_\mathrm{eff}+3 \frac{\dot R}{R} \left(\rho_\mathrm{eff}+p_\mathrm{eff}\right)=0.
\eq
Note that in the classical theory, $M_\mathrm{eff}$ equals $M$ and thus $p_\mathrm{eff}=0$ which is consistent with the fact that the matter content only consists of dust. However, in a model with quantum corrections, whether originating from just holonomies or holonomies and gauge-covariant fluxes, $M_\mathrm{eff}$ is in general time-dependent  and $p_\mathrm{eff}$ becomes nonzero. For this reason it is necessary to match the interior quantum modified spacetime with a generalized Vaidya spacetime. 

Finally, it is important to note that  the different effective masses in the collapsing and expanding phases does not imply the violation of the energy conservation law. The change in the effective mass is completely due to the quantum modifications in the gravitational sector of the Hamiltonian constraint. The matter sector of the Hamiltonian constraint remains untouched as can be seen by comparing the classical Hamiltonian (\ref{Hamiltonian}) and the effective Hamiltonian (\ref{ham2}). As a result, the energy density of the dust cloud still evolves according to $\rho=\frac{3M}{4\pi R^3}$ in both collapsing and expanding phases and the conservation law $\dot \rho+3H\rho=0$ always holds. Moreover, even in terms of the effective energy density and the effective pressure, the energy conservation law \eqref{energy conservation in effective quantities} holds for both branches.

\subsection{Numerical results of the effective dynamics}

We now present numerical results for the effective dynamics of the loop quantized dust outermost shell in the marginally bound case and compare the distinctive features resulting from the Hamiltonian (\ref{ham1}) and (\ref{ham2}). For convenience, the first model with only holonomy corrections  is called model A and the second model which is quantized by employing both holonomies and gauge covariant fluxes is called model B. Without any loss of generality, we set the initial time at $\tau=0$. The parameter space consists of the initial values of the radius $R_i$ and the dust mass $M$.  We find that  for a generic set of the initial conditions $(R_i, M)$, the singularity point at $R=0$ is replaced by a bounce in both models, but with important differences in physics of the bounce and post-bounce dynamics. Before the bounce, the dust cloud collapses in a contracting phase, and after the bounce, the dust cloud keeps expanding toward infinity. 

\begin{figure}
{
\includegraphics[width=8cm]{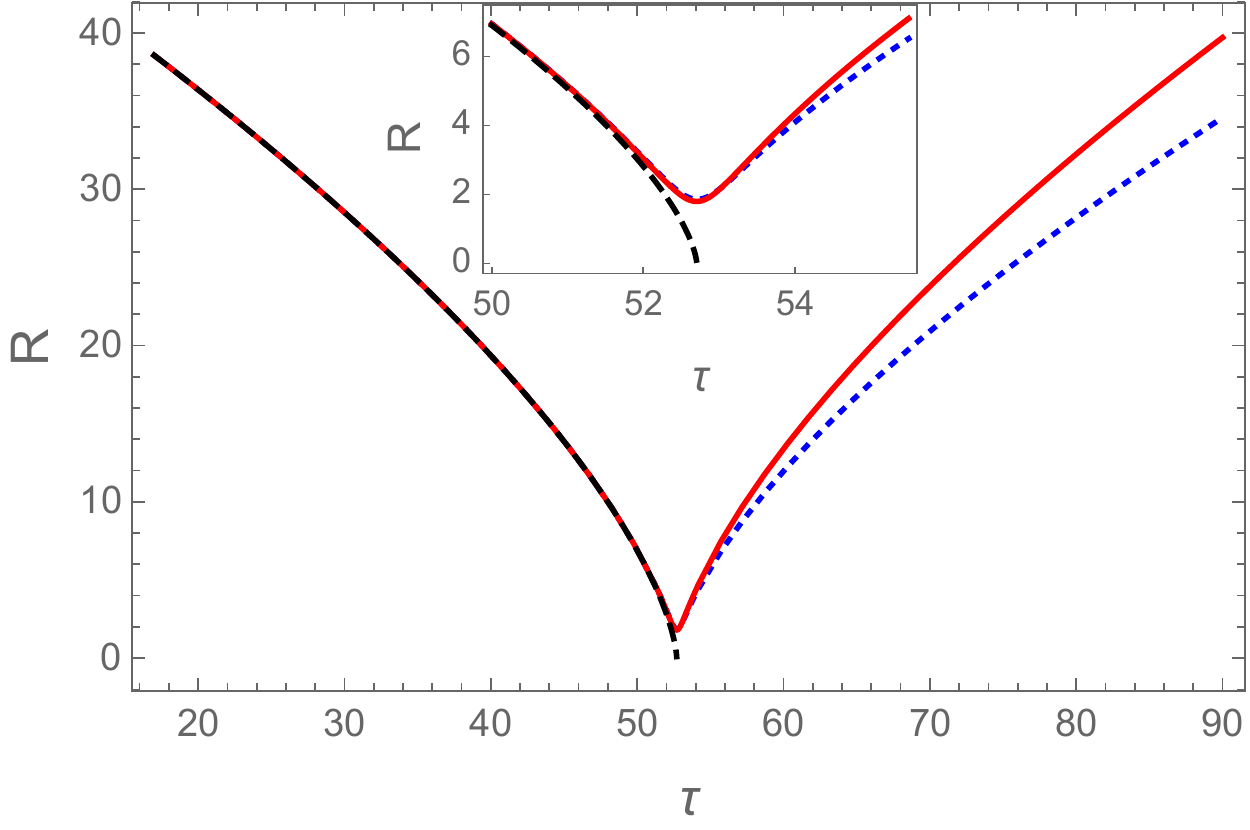}
\includegraphics[width=8.3cm]{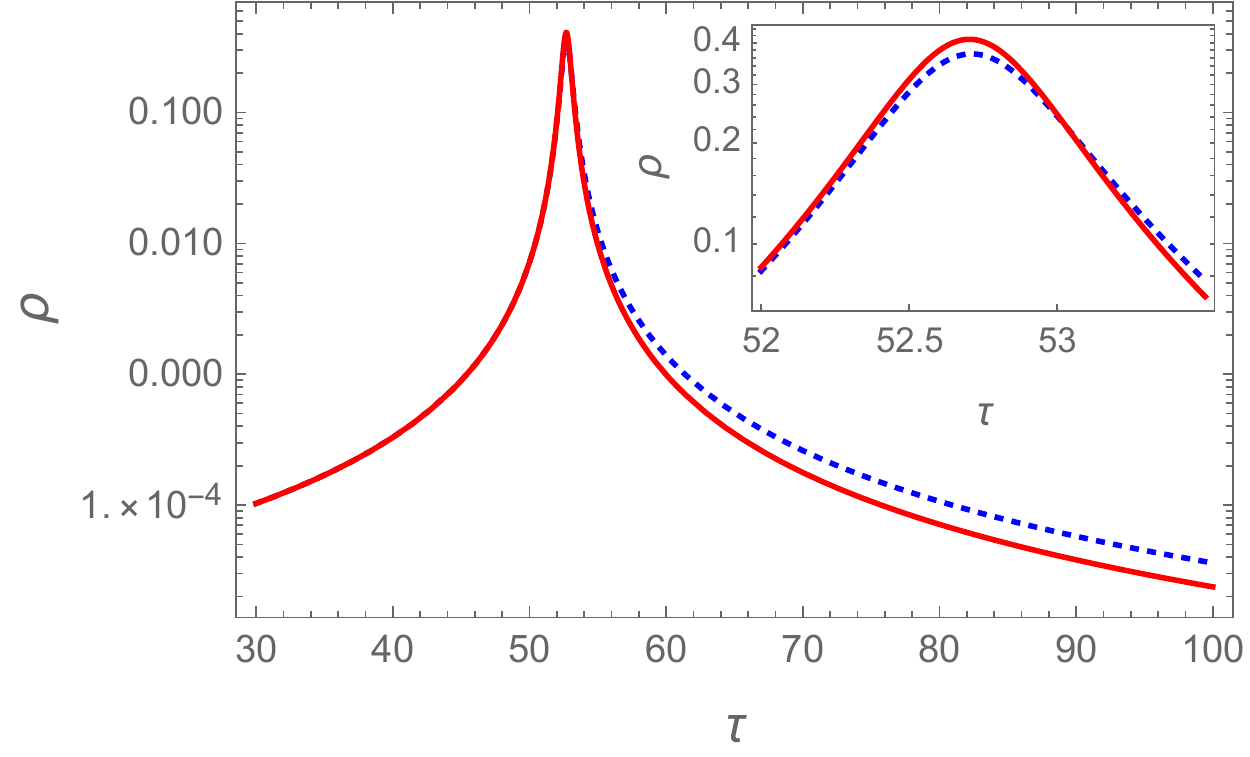}
}
\caption{With the initial conditions $R_i=50, M=10$, the classical  trajectory of $R$ (black dashed) is compared with those from the effective dynamics prescribed by holonomy-triad quantization (\ref{ham1}) (red solid) and holonomy-gauge-covariant flux quantization (\ref{ham2}) (blue dotted) in the left panel. In the inset plot, it can be seen that the classical trajectory (black dashed curve) ends at the singularity point when $R=0$ while in the other two trajectories, a bounce occurs at nonzero radius. In the right panel, we can find the difference in the energy densities of two loop quantized models mainly in the expanding branch. The inset plot shows the difference of the maximum energy densities in the two models.  All values are in Planck units.}
\label{radiusandrho}
\end{figure}

In Fig. \ref{radiusandrho}, we show a representative case with $R_i=50$ and $M=10$ (in Planck units). The left panel depicts the evolution of the radius $R$ of the shell in the classical theory (black dashed curve), in model A with holonomy corrections (red solid curve) and in  model B with both holonomies and gauge covariant fluxes (blue dotted curve). In the classical theory, the radius becomes zero at a finite proper time while in models A and B,  the radius of the shell experiences a bounce at the maximum energy density and the dust shell enters into an expanding phase afterwards. The bounce in both models takes place at  around $\tau\approx52.70$ (in Planck units). The difference between two loop quantized models mainly lies in the expanding phase and they also have different maximum energy densities. In model A, the evolution is symmetric with respect to the bounce while in model B, we observe an asymmetric bounce due to  the gauge covariant fluxes. Starting with the same initial conditions, the expansion rate of the dust cloud in model B  is smaller than that in model A. Besides, the maximum energy density in model A is $\rho^{\scriptscriptstyle{\mathrm{hol}}}_\mathrm{max}\approx0.409$ while in model B, the maximum energy density is $\rho^\mathrm{g.c.}_\mathrm{max}\approx0.370$ in Planck units.

\begin{figure}
{
\includegraphics[width=8cm]{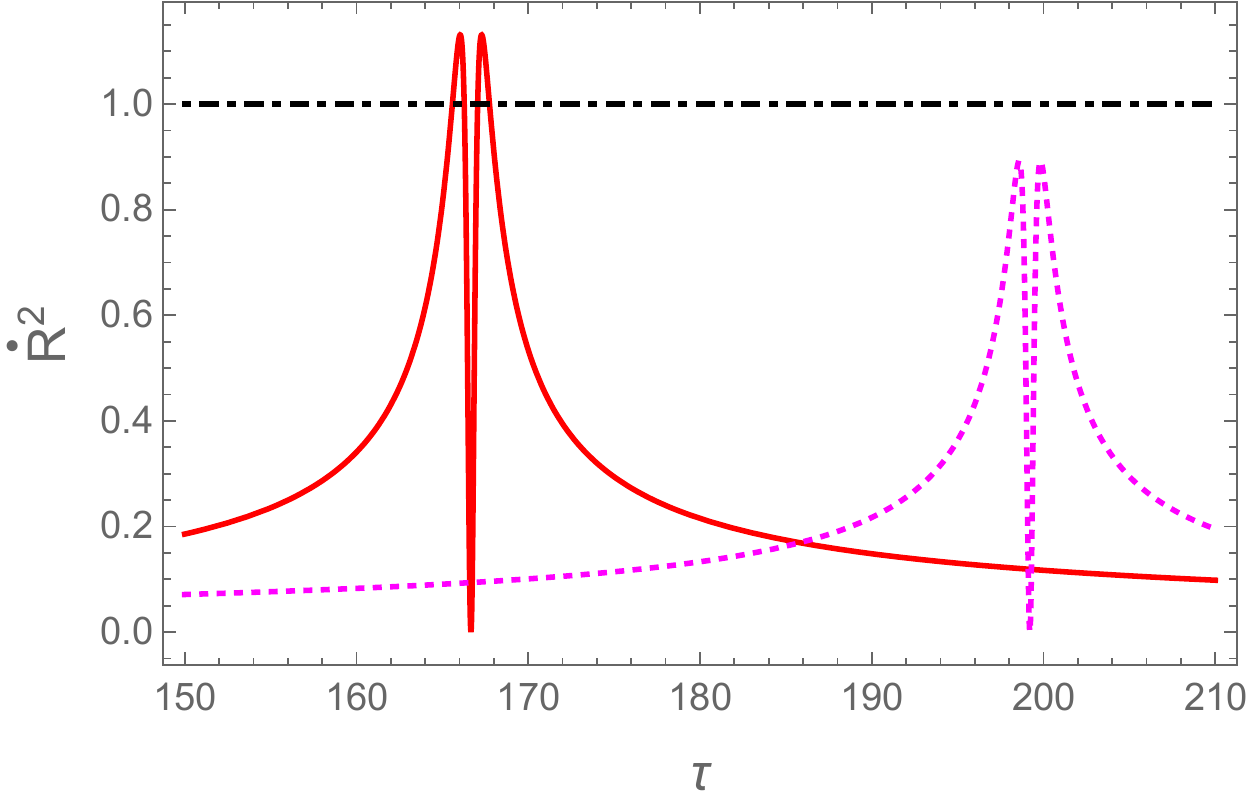}
\includegraphics[width=8cm]{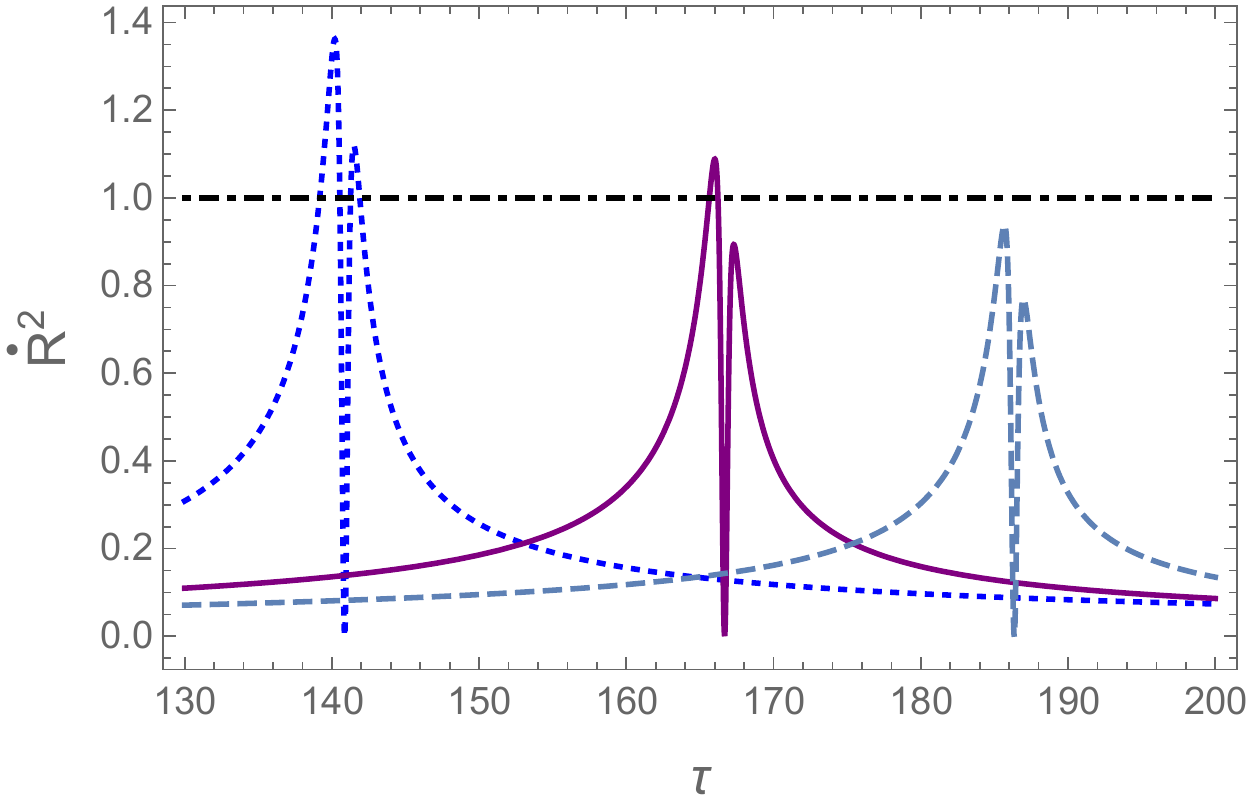}
}
\caption{With the fixed initial radius $R_i=50$, we change the dust mass $M$ in both of the models to show the effects of the dust mass on the formation of the trapped surfaces corresponding to black hole (white hole) before (after) the bounce point. In the left panel, the dust mass is set to $M=1.0$ (red solid) and $M=0.7$ (magenta dotted) in the first model with only holonomy corrections. In the right panel, the dust mass  from left to right is set, respectively,  to  $M=1.4$ (blue dotted), $M=1.0$ (purple solid) and $M=0.8$ (dashed) in the second model which also considers the gauge covariant flux. The trapped horizon forms when $\dot R^2\ge1$.}
\label{velocity}
\end{figure}

We find for both models, the formation of a trapped surface does not depend on the initial values of the radius which would only affect the bounce time. The only parameter that affects the formation of the trapped surface is the dust mass $M$. Moreover, we observe different patterns in two loop quantized models which are summarized in Fig. \ref{velocity}. In the figure, we present the $\dot R^2$ plot  for both models. The left panel is for  model A and the right panel is for model B. As discussed in the last subsection, in the contracting phase when the dust cloud collapses, the black hole forms in the period when $\dot R^2>1$. On the other hand, a white hole would form if $\dot R^2$ becomes larger than unity again in the expanding phase after the bounce. In model A (left panel), the peaks of $\dot R^2$ are symmetric with respect to the bounce. This shows that if the contrating branch produces a black hole then the expanding branch produces its twin white hole.  There is a threshold value of the dust mass which determines whether black hole or white hole would form. Its analytic value $M^*\approx0.831$ given in (\ref{threshold}) exactly matches with our numerical results. In particular, our numerical investigations for masses below this threshold did not find formation of a trapped surface. In the figure, we show two cases with $M=1.0$ and $M=0.7$. In the former case, a black (white) hole forms near the bounce  while in the latter case, neither a black hole nor a white hole would form during the entire evolution. Finally, once the dust mass is fixed,  from Eq. (\ref{velocity-squared}) one can deduce at what energy density the trapped surfaces would form or vanish. In the case  of $M=1.0$, we find that during the collapse of the dust cloud, the black hole forms at $\rho=0.041$ and evaporates at $\rho=0.188$. Owing to the symmetric bounce, when the dust cloud enters into the expanding phase after the bounce, the white hole would form at $\rho=0.188$ and vanish at $\rho=0.041$. Increasing the dust mass would decrease/increase the energy densities at which the trapped surfaces form/vanish during the collapsing phase of the dust cloud. For example, when $M=2000$, the black hole would form at $\rho=7.460\times10^{-9}$ and then evaporate at $\rho=0.408$. As $\dot R^2$ changes monotonically with the dust mass in (\ref{velocity-squared}), the energy density at which the trapped surfaces form or vanish also changes monotonically with the dust mass. 

The situation becomes  richer for model B (right panel in Fig. 2) where we find that $\dot R^2$ is asymmetric with respect to the bounce point. As a result, there are two characteristic dust masses $M_1=0.880$ and $M_2=1.184$. When the dust mass $M<M_1$, there would be no black hole or white hole as depicted by the rightmost  dashed curve in the right panel. When $M_1<M<M_2$, only the black hole can form during the collapse of the dust cloud, there would be no white hole in the expanding phase. This case corresponds to the one depicted by the middle purple solid curve in the right panel. Lastly, if $M>M_2$, both a black hole and a white hole can form as depicted by the blue dotdashed curve. Since the closed forms of the dynamical equations which yield an analytical value of mass threshold are not available for this model, the above threshold values of the dust mass for the formation of the black hole and white hole are determined numerically. We have checked with various different initial radii and find the same threshold values of the dust mass. Moreover, given a specific dust mass, one can only find the energy densities at which the black hole and the white hole would form and vanish through the numerical solutions.  In the case mentioned in the right panel of Fig. \ref{velocity}, we find for $M=1.4$, the black hole would form at $\rho=0.0126$ and evaporate at $\rho=0.154$. Correspondingly, the white hole would form at $\rho=0.130$ and vanish at $\rho=0.0364$. One can also increase the dust mass and find the similar patterns as in model A with only holonomy corrections. For example, when $M=2000$, the black hole would form at $\rho=2.98\times10^{-8}$ and evaporate at $\rho = 0.368534$ while the corresponding white hole would form at $\rho \approx 0.368517$ and vanish at $\rho=1.16\times10^{-7}$.  We find that the formation of trapped surfaces does not depend on the initial radius.  Note that there also exists an asymmetry in the energy densities for the formation or the evaporation of the black hole and the white hole in model B. Thus we find a key difference between the physics of model A and B. Contrary to the model where triads and holonomies are used,  in presence of gauge-covariant flux modifications a black hole-white hole twin system is not possible and there can be situations where only a black hole forms.

\begin{figure}
{
\includegraphics[width=8cm]{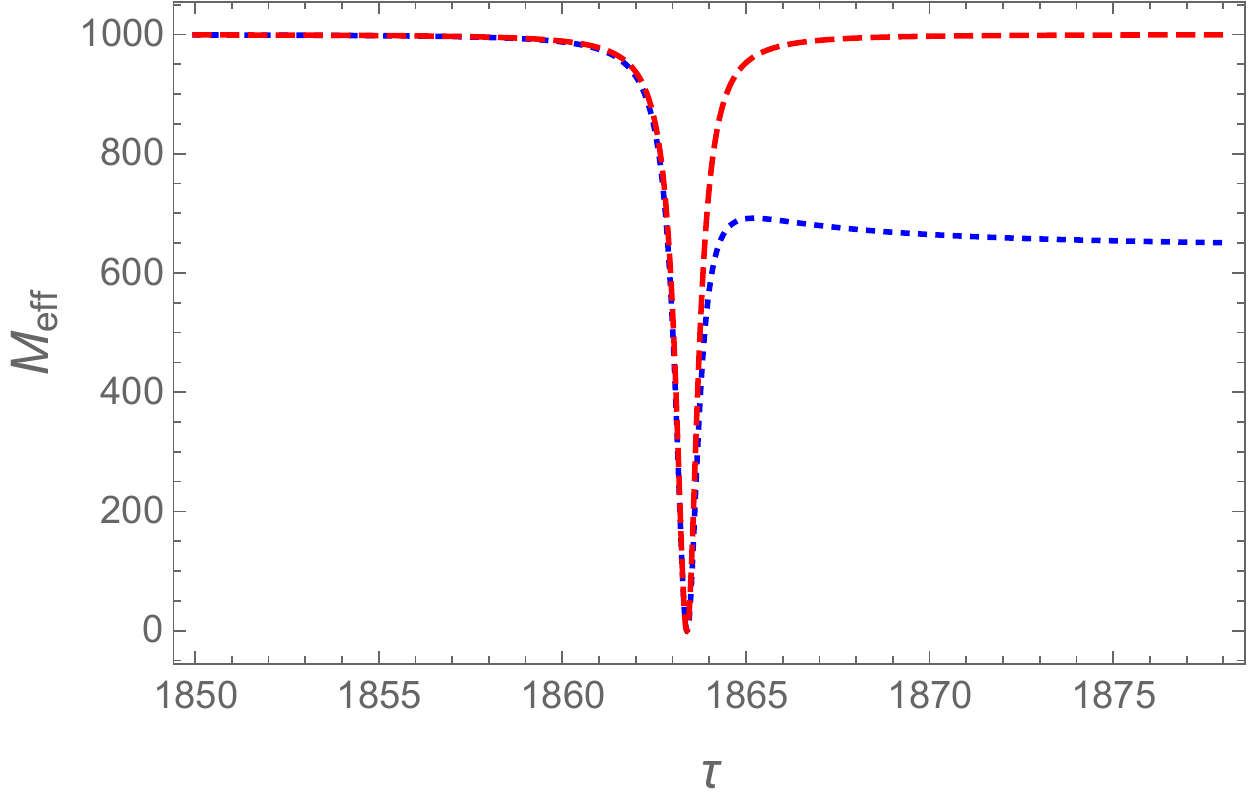}
}
\caption{With the initial conditions $R_i=2500, M=1000$, the effective mass in two models are compared near the bounce.  The difference between two models exhibits itself in the expanding phase where the red dashed  curve returns to its initial  value quickly after the bounce while the blue dotted curve can only return to  $2/\pi$ of its original value.}
\label{effective mass}
\end{figure} 

In addition, we also find the difference in the effective masses between two models. In Fig. \ref{effective mass}, we choose the initial conditions $R_i=2500, M=1000$ (in Planck units) so that both black hole and the white hole will form in the two considered models. From the figure, one can find initially when the quantum gravity effects can be ignored, the effective mass is equal to the dust mass in both cases. As the dust cloud continues to collapse, the effective mass drops slowly initially. It is only near the bounce that the effective mass starts to change drastically.  Right at the bounce, the effective mass vanishes which implies, for an exterior observer, the spacetime reduces to the flat Minkowski spacetime at the bounce. The remarkable difference between two models takes place in the expanding phase. In model A (red dashed curve), the dust cloud can return to its initial configuration with a reversed velocity, as the quantum gravity effects disappear, the effective mass also returns to its initial value. However, in model B with modifications from the gauge covariant fluxes, the effective mass can only at most return to about $\alpha M\approx636.62$ which is exactly given by our analytical predictions in \eqref{mass in expanding phase}. Finally, since the $M_\mathrm{eff}$ is asymmetric with respect to the bounce in model B, the masses of the black hole and the white hole also evolve asymmetrically with respect to the bounce.

\begin{figure}
{
\includegraphics[width=8cm]{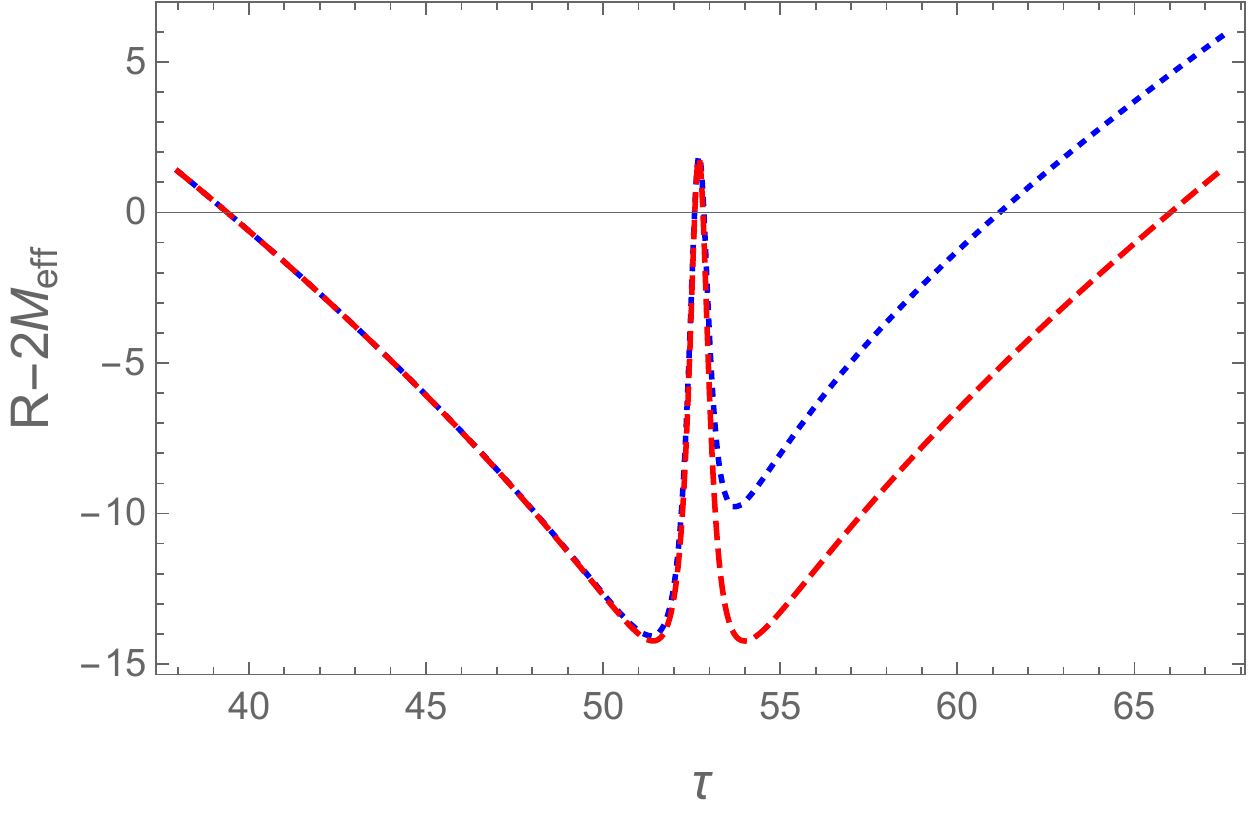}
}
\caption{With the initial conditions $R_i=50, M=10$,  we show explicitly the evolution of the trapped horizon near the bounce.  The bounce takes place at the peak in the middle of the plot. The red dashed curve is for  model  A with only the holonomy corrections while the blue dotted curve is for model  B which also considers gauge covariant fluxes. }
\label{rminus2m}
\end{figure}

\begin{figure}
{
\includegraphics[width=8cm]{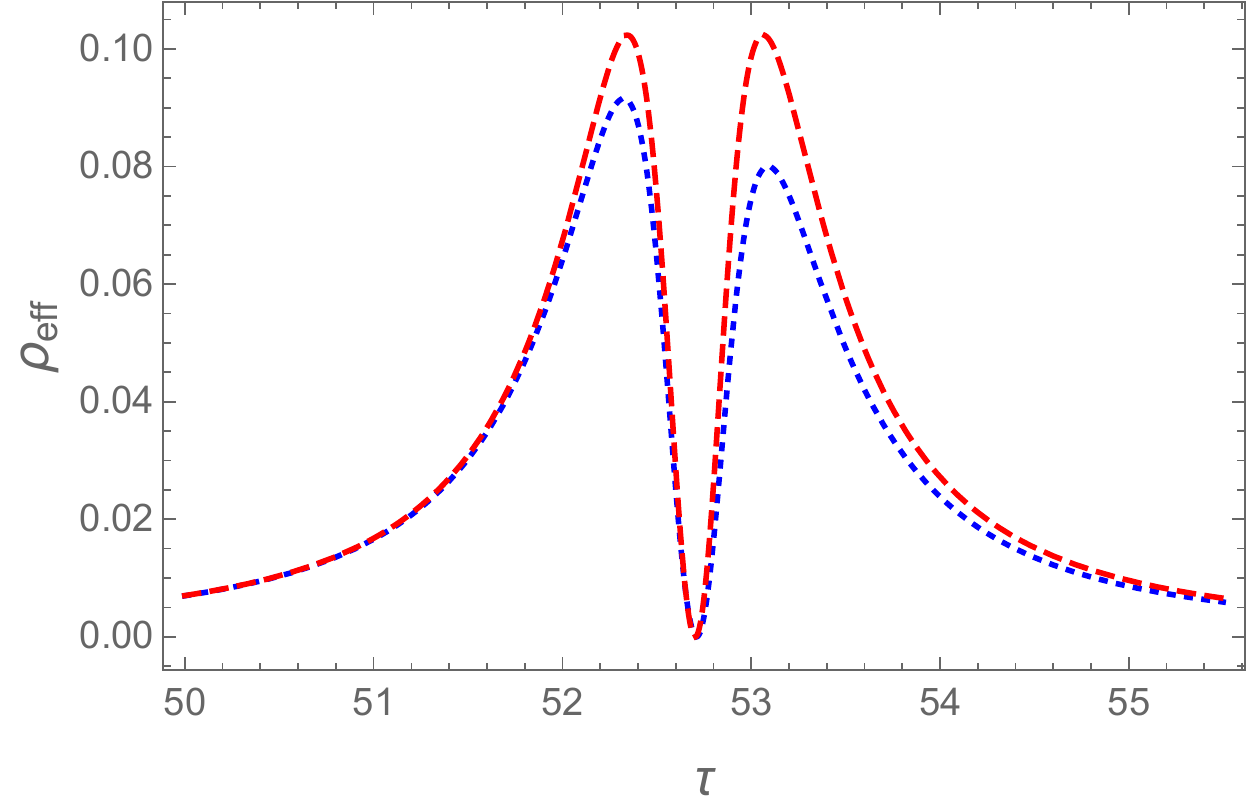}
\includegraphics[width=8cm]{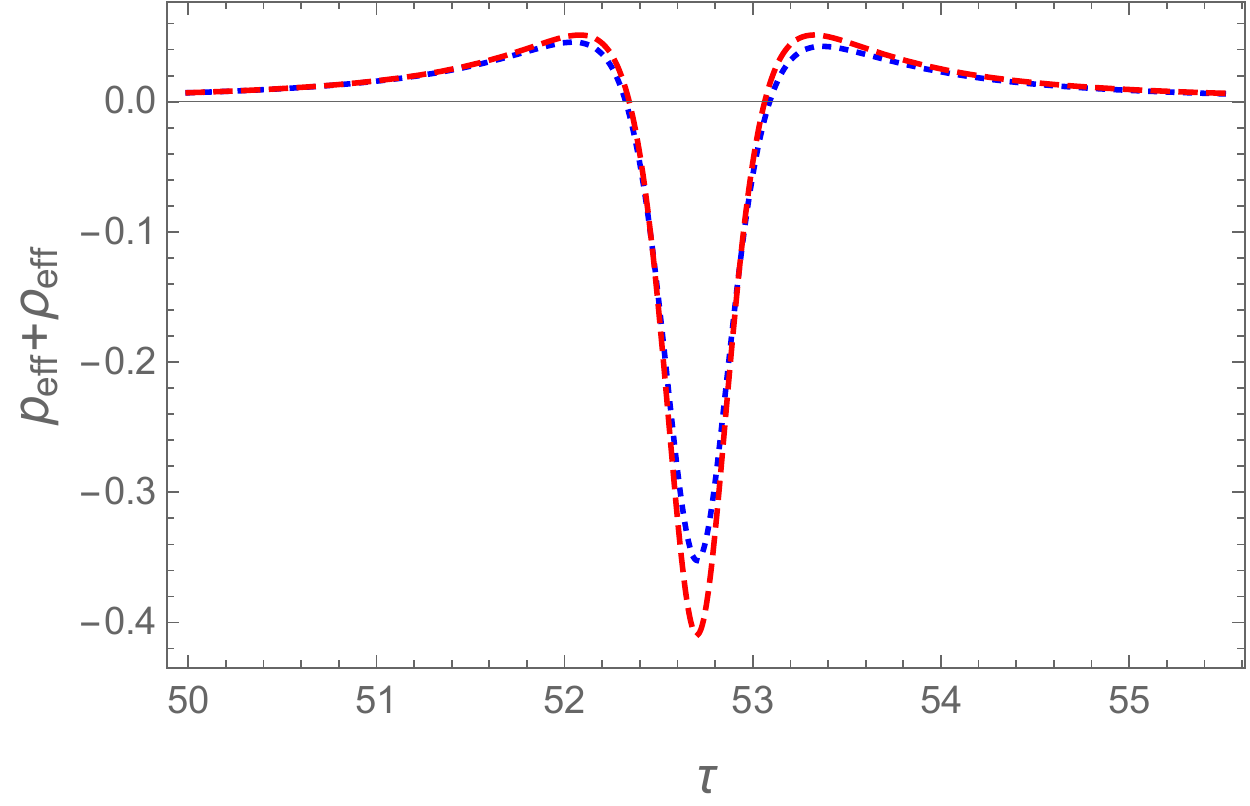}
}
\caption{With the initial conditions $R_i=50, M=10$,  apart from some quantitative differences,  the effective energy density in  model A (red dashed curves) and model B (blue dotted curves) have the similar behavior. In particular, the effective energy densities vanish at the bounce and the weak energy condition is violated in a neighborhood  of the bounce.}
\label{weak energy condition}
\end{figure} 

As the effective mass $M_\mathrm{eff}$ equals the mass of the black hole and white hole once the trapped surface is formed, one can also track the formation of the trapped surface using the $(R-2M_\mathrm{eff})$ plot. We give  such an example in Fig. \ref{rminus2m} with the initial conditions $R_i=50, M=10$. The black hole and the white hole are formed when $R<2M_\mathrm{eff}$. The central maximum at around $t\approx52.70$ in the plot corresponds to the bounce point. From the figure, we can clearly see when the trapped surface forms and disappears in both of the models. When the dust cloud is in the collapsing phase, the radius decreases faster than the effective mass. The trapped surface forms in the regime $39.39<t<52.58$ for which $R<2M_\mathrm{eff}$ and disappears in the interval $52.58<t<58.84$ when $R>2M_\mathrm{eff}$. Right at the bounce $t\approx52.70$, $M_\mathrm{eff}$ vanishes and $(R-2M_\mathrm{eff})$ attains its local maximum.  After the bounce, in order to form a trapped surface during the expanding phase, $2M_\mathrm{eff}$ which equals zero at the bounce has to increase more quickly than $R$ so that  $(R-2M_\mathrm{eff})$ becomes negative again at some instant of time (in the current case, this time is $t\approx58.84$ for both models.  In this expanding branch, the difference between the two models exhibits itself in terms of the duration of the trapped surfaces. To be specific, in model A (red dashed curve) which includes only  holonomy corrections, the trapped surface forms at around $t\approx58.84$ and disappears at around $t\approx 66.02$ while in model B (blue dotted curve) which includes quantum effects from holonomy corrections and the gauge covariant fluxes, the trapped surface lasts from $t \approx 58.84$ to $t\approx 61.19$. As a result,  in model A,  the black hole and white hole have the same ``lifetime" in terms of  the proper time $\tau$, while in model B, the black hole outlives the white hole in the proper time $\tau$. 

While there are qualitative differences between the two models, the behavior of the effective energy density and the violation of the weak energy conditions when measured with respect to effective energy densities in the two models is similar. In Fig. \ref{weak energy condition}, we present a typical evolution of the effective energy density  in both models with the initial conditions $R_i=50, M=10$. One can find two models share the similar behavior: as the dust cloud collapses, the effective energy density initially increases until its maximum value before the bounce then it decreases rapidly. At the bounce, the effective energy density vanishes which implies that an outside observer can only see an asymptotic flat Minkowski  spacetime at the bounce. Afterwards, the effective energy densities increases again to its maximum  as the dust cloud starts to expand. This results in the formation of the white hole in the expanding phase when $R<2M_\mathrm{eff}$.  As the radius keeps increasing, the effective energy density would finally decrease. Finally, in the right panel, we can find near the bounce between $\tau\approx 52.32$ and $\tau\approx 53.09$, the weak energy condition in terms of the effective density and pressure is violated in both models.

It is important to note that the above qualitative differences and similarities between the two models are robust with respect to the change in the dust mass and the initial radius of the dust cloud. In particular, the symmetric bounce in model A and asymmetric bounce in model B, the threshold values of the dust mass for the formation of the trapped surfaces in both models, as well as the black hole and white hole symmetry (asymmetry) in model A (B)  are the properties of models independent of the specific choices of the initial conditions and were found to hold for a large range of values of mass parameter.

\section{Conclusions}
\label{sec:summary}

In this manuscript, we have studied and compared the dynamical consequences of two loop quantizations of the dust shell model. The first quantization uses the holonomies and triads while the second quantization employs the holonomies and the gauge covariant fluxes. Note that in standard LQC the use of triads follows because of the symmetry reduction and homogeneity. On the other hand the motivation to use gauge-covariant fluxes arises from treating holonomies and fluxes at a similar level during quantization in LQG \cite{Thiemann_2001} and to obtain the effective Hamiltonian in LQC
as an expectation value of the scalar constraint operator using suitable coherent states in LQG. 
In the classical dust shell model, we assume the LTB dust spacetime with marginally bound condition so that the initial dust velocity  at spatial infinity vanishes. Moreover, we also assume a homogeneous dust density so that all shells of the dust cloud collapse and expand at the same rate. In this way, there is no shell crossing singularity and all of the shells reach the central singularity at $R=0$ simultaneously. In both of the loop quantized models, the said central singularity is resolved for a generic set of the initial conditions for the collapsing dust cloud. The dust shell  bounces back after some period of collapse when the maximum energy density allowed in each model is reached. Afterwards, the dust cloud keeps expanding until all the matter is radiated to infinity. 

Besides the generic resolution of the central singularity, in both models, the formation of the trapped surfaces only depends on the dust mass $M$.  However, there are important qualitative distinctions in the dynamics of two loop quantized models.  In the first model with only holonomy corrections, the evolution of the radius, the velocity of the dust shell, the effective mass and the effective energy density is symmetric with respect to the bounce point, while in the second model with both holonomy corrections and modifications from the gauge covariant fluxes, the evolution of the corresponding variables become asymmetric with respect to the bounce. The symmetric/asymmetric bounce results in the following physical consequences.  In the  first model, we find a threshold value of the dust mass $M^*$ below which no trapped surface would form during the entire evolution of the dust cloud. It is only when the dust mass is larger than $M^*$ then the trapped surfaces could form on both sides of the bounce. The trapped surface which forms during the collapse of the dust cloud corresponds to a dynamical black hole while the trapped surface which forms during the expansion of the dust cloud after the bounce is a dynamical white hole. The black hole and the white hole lie symmetrically on both sides of the  bounce point. Between them is an asymptotic flat Minkowski spacetime,  owing to the vanishing effective mass (effective energy density) at the bounce point.  

On the other hand, in the second model which includes gauge-covariant flux modifications, we find a much richer situation. In contrast to the first model, the formation of the black hole in the collapsing phase does not guarantee the formation of the white hole in the expanding phase. There actually exist two characteristic dust masses $M_1$ and $M_2$ ($M_1<M_2$). When the mass of shell is less than  $M_1$, neither black hole nor white hole would form during the collapse or the expansion of the dust cloud. When the dust mass lies between $M_1$ and $M_2$, only the black hole can form during the collapse of the dust cloud. Finally,  when the  dust mass is larger than $M_2$, both black hole and white hole can form on different sides of the bounce point. For the last case,  unlike in the first model, the evolution of the black and the white hole  is not symmetric with respect to the bounce point. 

 Another remarkable difference between two loop quantized models lies in the behavior of the effective mass and the duration of the trapped surfaces in two models. In the first model with only holonomy corrections, the effective mass evolves symmetrically with respect to the bounce. It  tends to the same dust mass on both sides of the bounce when the dust energy density approaches zero.  However, in the second model, the asymptotic value of the effective mass of the dust cloud in the expanding phase is only  $2/\pi$ of its initial value in the collapsing phase. As the effective mass remains almost constant in most of the times of the evolution and it only changes drastically near the bounce, this asymmetry can also be interpreted as the asymmetry in the masses of the black hole and white hole formed during the collapse and the expanding phases respectively. Moreover, in the first model, the black hole and the white hole have the same lifetime in terms of the proper time  $\tau$ while in the second model, the black hole outlives the white hole as a consequence of the asymmetric bounce which is a unique feature of the second model. In summary, our results show that various aspects of the black hole-white hole symmetry which exist in models based on holonomy modifications are non-existent when one also includes gauge-covariant flux modifications motivated by loop quantum gravity. Further, our analysis shows that even for the simplest situation of the marginally bound case different quantization prescriptions can result in qualitatively different physics for the white hole spacetime.

\section*{Acknowledgements} 
We thank Anzhong Wang and  Jahnvi Verma  for comments on the manuscript. This work is supported by the DFG-NSF grants PHY-1912274 and 425333893, and PHY-1454832.


\end{document}